\documentclass[a4paper,fleqn]{cas-sc}

\usepackage[numbers]{natbib}
\def\tsc#1{\csdef{#1}{\textsc{\lowercase{#1}}\xspace}}
\tsc{WGM}
\tsc{QE}
\usepackage{gensymb}

\begin{document}
\let\WriteBookmarks\relax
\def\floatpagepagefraction{1}
\def\textpagefraction{.001}

\shorttitle{Ambiguity in B-Site cation ordering: A Case study of the double perovskite Ca$_{2}$CoNbO$_{6}$}    

\shortauthors{S.A. Artiukova et~al.}  

\title [mode = title]{Ambiguity in B-Site cation ordering: A Case study of the double perovskite Ca$_{2}$CoNbO$_{6}$}  



%
\author[inst6]{Svetlana A. Artiukova}[orcid=0009-0007-6764-7326]
\ead{svetlana.artiukova@skoltech.ru}
\credit{Investigation, Visualization, Software, Validation, Writing – original draft}
\affiliation[inst6]{organization={Skolkovo Institute of Science and Technology}, addressline={Skolkovo Innovation Center, Bolshoy Boulevard, 30, bld. 1},
city={Moscow},
postcode={121205},
country={Russia}}
\author[inst3]{Ivan V. Yatsyk}
\ead{i.yatzyk@gmail.com}
\credit{ Investigation, Visualization}
\affiliation[inst3]{organization={Zavoisky Physical-Technical Institute, Federal Research Center “Kazan Scientific Center of RAS”},
addressline={Sibirsky tract, 10/7}, 
city={Kazan},
postcode={420029}, 
country={Russia}}
\author[inst2]{Ruslan G. Batulin}
\ead{tokamak@yandex.ru}
\credit{Investigation}
\affiliation[inst2]{organization={Kazan (Volga Region) Federal University},
addressline={Kremlevskaya st., 18}, 
city={Kazan},
postcode={420008}, 
country={Russia}}
\author[inst4]{Yulia A. Deeva} 
\ead{juliahik@mail.ru}
\credit{Investigation}
\affiliation[inst4]{organization={Institute of Solid State Chemistry of the Russian Academy of Sciences (UB)}, addressline={Pervomaiskaya St., 91},
city={Ekaterinburg},
postcode={620990},
country={Russia}}
\author[inst4]{Tatiana I. Chupakhina}
\ead{chupakhina@yandex.ru}
\credit{Investigation,  methodology}
\author[inst4]{Vladislav V. Bazhal}
\ead{bazhal70780@gmail.com}
\credit{Investigation,  methodology}
\author[inst4]{Alexandr I. Balitskiy}
\ead{bal.tot.sia@gmail.com}
\credit{Investigation,  methodology}
\author[inst3]{Rushana M. Eremina}
\ead{reremina@yandex.ru}
\credit{Conceptualization, Project administration, Supervision, Writing - original draft, Visualization, Resources}
\author[inst3,inst5]{Dina I. Fazlizhanova}
\cormark[1]
\ead{dina.fazlik@gmail.com}
\credit{Formal analysis, Validation, Investigation, Writing - original graft, Visualization}
\affiliation[inst5]{organization={Physics Department, National University of Science and Technology “MISIS”},
addressline={Leninskiy Prospekt, 4}, 
city={Moscow},
postcode={119049}, 
country={Russia}}













\cortext[1]{Corresponding author}



\begin{abstract}
The ordering of cations in the B sublattice remains a challenging issue in double perovskites. In this work, a combined experimental and theoretical approach was employed to investigate the Co/Nb distribution in Ca$_{2}$CoNbO$_{6}$ and its influence on magnetic and transport properties. The density functional theory, supported by magnetic susceptibility measurements, indicates that Co adopts a high-spin Co$^{3+}$ state. No long-range magnetic ordering was observed down to low temperatures; however, the presence of short-range correlations points to the partial disorder in the Co/Nb sublattice. This interpretation is further supported by electron paramagnetic resonance, which also reveals slight oxygen nonstoichiometry.

 Electrical transport follows a small-polaron hopping mechanism with an activation energy of 0.25 eV. The Seebeck coefficient reaches 0.4 mV/K at 600 K.
\end{abstract}


\begin{highlights}
\item Predominantly rock-salt Co/Nb distribution at the B-site with an approximate ratio of 0.65/0.35.
\item High-spin cobalt state confirmed by combined magnetic and computational data
\item Magnetic measurements suggest that Co$^{2+}$ predominantly occupies Nb–O–Co–O–Nb local environments
\item Small polaron hopping governs electrical transport with low activation energy of 0.25 eV
\item Ca$_2$CoNbO$_6$ demonstrates large positive Seebeck coefficient of 0.4 mV/K at 600 K 
\end{highlights}


\maketitle

\section{Introduction}
\label{sec1}
Double perovskite oxides with the general formula $A_2B'B''O_6$, where the A-site cation is a larger ion, typically an alkaline-earth metal or rare-earth element that stabilizes the framework, while the B-site cations are smaller transition metals or elements of the main-group that occupy octahedral sites and exhibit various oxidation states, have attracted considerable attention over the past few decades due to their diverse and tunable physical properties \cite{VASALA20151}. Extensive research has explored their structural, magnetic, and electronic characteristics, revealing a wide spectrum of  phenomena, including magnetoresistance \cite{kobayashi1998room, mahato2010colossal}, multiferroicity \cite{du2010magnetic,liu2022multiferroic,yanez2011multiferroic}, and anomalous Hall effects \cite{chakraborty2022berry}. Moreover, double perovskites display various electronic behaviors such as half-metallicity \cite{D2TC03199J}, metal-insulator transitions \cite{poddar2004metal,streltsov2022ground,chen2018magnetically}, and both n-type and p-type conduction mechanisms, some compounds exhibiting high Seebeck coefficients \cite{VASALA20151}. They also host a rich variety of magnetic ground states, including ferromagnetism, antiferromagnetism, ferrimagnetism, spin-glass behavior, and other exotic phases. Many of these properties are highly sensitive to the degree of B-site cation ordering, making the control of cation distribution a central strategy for tuning their functional behavior \cite{Karppinen2005}. These compounds can exhibit various types of B-site ordering, including the common rock-salt arrangement, as well as less frequent layered and columnar patterns \cite{VASALA20151}. In one of the most studied double perovskite Sr$_2$FeMoO$_6$, the degree of B-site cation order significantly affects ferromagnetic properties, with saturation magnetization decreasing linearly with increased disorder \cite{balcells2001cationic}. Proper synthesis methods can achieve high degrees of order and consequently high magnetization \cite{huang2004simple}. Conversely, ordered ruthenate compounds Ln$_2$LiRuO$_6$, where (Ln = Pr, Nd, Sm, Eu, Gd, and Tb) are predominantly antiferromagnetic \cite{makowski2009coupled}, while quenched samples with B-site disorder of La$_2$NiMnO$_6$ exhibit frustrated magnetic behavior \cite{dass2003oxygen}. Many compounds demonstrate spin-glass-like behavior due to notable cation disorder \cite{battle1989spin, battle1995sol}, as seen in disordered Sr\(_2\)Fe\(B''\)O\(_6\) compounds with \(B''\)= (Nb, Ta, Ru) \cite{battle1989spin, battle1995investigation,kashima2002low}. 

Transport properties can also be influenced by disorder at the B-site. For instance, in Sr$_2$FeMoO$_6$ a higher degree of order generally corresponds to lower resistance, and the transition from polaronic to itinerant behavior occurs around an order parameter $\xi = 0.90$ \cite{huang2006systematic}. Moreover, cation disorder tends to disrupt the half-metallic nature of this and similar compounds, thereby reducing the tunneling magnetoresistance effect \cite{sarma2000magnetoresistance,garcia2001finding}.

The degree of B-site cation ordering in double perovskites is governed by several key factors, primarily the charge difference ($\Delta Z_B $) and ionic radius difference ($\Delta r_B$) between the B-site cations. In general, larger values of $\Delta Z_B$ and $\Delta r_B$ promote ordering by reducing electrostatic repulsion and lattice strain, respectively \cite{ANDERSON1993197,B926757C}. For example, systems with $\Delta Z_B > 2$ typically favor fully ordered structures. In contrast, when $\Delta Z_B = 2$, the degree of ordering becomes more sensitive to the size of mismatch: for compounds with valence configurations $A^{2+}_2 B'^{3+} B''^{5+} O_6$ and small ionic radius differences ($\Delta r_B < 0.2$ \AA), a wide range of partial ordering phenomena was observed, from nearly disordered to fully ordered states. The general trend is that smaller ($\Delta r_B$) tends to correlate with lower degrees of ordering \cite{VASALA20151}.
 
One of the key factors influencing ordering in the structures with $Delta Z_B=2$ is the synthesis temperature. For instance, it was shown that for Sr$_2$AlBO$_6$ (B=Nb, Ta) higher temperatures and longer annealing times generally promote ordering through thermally activated cation diffusion \cite{woodward1994order}. The pressure during synthesis also plays a significant role; reducing the lattice volume via external pressure favors ordered structures by minimizing electrostatic repulsion between highly charged B-site cations. Additionally, subtle variations in synthesis conditions, such as cooling rates, can influence oxidation states and indirectly affect ordering \cite{VASALA20151}. Understanding these ordering phenomena is essential for tuning the properties of these materials for various applications. 
  
The synthesis of the double perovskite Ca$_2$CoNbO$_6$ was originally reported by R. Shaheen and J.Bashir in \cite{Shaheen}. It was prepared by solid-state reaction, while its crystal structure has been refined using powder X-ray diffraction data in. The compound is monoclinically distorted and adopts the space group P2$_{1}$/n with the cell parameters  a = 5.4797(1) \AA, b = 5.6051(1) \AA, c = 7.8119(2) \AA, $\beta$  = 89.96(1)\degree.  Co and Nb are found to be distributed over the six coordinated octahedral sites in a rock salt arrangement. The refined Co–O and Nb–O bond lengths are 1.9788(2) \AA \
and 2.0642(2) \AA, respectively.  Replacement of niobium with tantalum leads to a change in the parameters of the crystal structure.

In this work, the double perovskite Ca$_2$CoNbO$_6$ was studied in order to investigate the B-site cation ordering and its impact on the material's structural, magnetic, and transport properties. The compound was synthesized using the pyrolysis method of nitrate-organic mixtures, followed by detailed characterization through X-ray diffraction (XRD), magnetization measurements, transport measurements, and electron spin resonance (ESR) spectroscopy. To complement the experimental findings, DFT calculations with Hubbard U correction were performed to analyze two possible distributions of Co and Nb - a rock-salt-type and alternating layers of the cations in order to evaluate their thermodynamic stability and magnetic configurations.

\section{Methods}
\subsection{Synthesis}
The double perovskite \( \text{Ca}_2\text{CoNbO}_6 \) was synthesized using the pyrolysis method of nitrate-organic mixtures of the corresponding components. The initial reagents included calcium carbonate (\(\text{CaCO}_3\)), cobalt(II) nitrate hexahydrate (\(\text{Co(NO}_3)_2\) $\cdot$ 6 H$_2$O), niobium(V) oxide (\(\text{Nb}_2\text{O}_5\)), and xylitol as an organic additive. The stoichiometric amount of \( \text{CaCO}_3 \) was dissolved in dilute nitric acid (1:1) and Co(NO$_3)_2$ $\cdot$ 6 H$_2$O was dissolved in distilled water. The resulting solutions were mixed, followed by the addition of a stoichiometric amount of \( \text{Nb}_2\text{O}_5 \). Ammonium hydroxide (\( \text{NH}_4\text{OH} \)) was then introduced to the mixture to adjust the pH to 12, and the solution was left to stand for 24 hours. The reaction mixture was subsequently heated to \( 300-350^\circ\text{C} \) until ignition occurred. The resulting nanodispersed powder was calcined at \( 950^\circ\text{C} \) for 4 hours to remove carbon. The powder was then pressed into tablets and annealed stepwise, with intermediate grinding and repressing, at temperatures of \( 1000^\circ\text{C} \), \( 1050^\circ\text{C} \), \( 1100^\circ\text{C} \), and \( 1200^\circ\text{C} \) for 8 hours each. 
\subsection{Sample characterization}
X-ray diffraction (XRD) measurements were conducted using the Shimadzu XRD-7000 S automatic diffractometer, with an exposure time of 3–5 seconds per point. The X-ray pattern analysis was carried out using the FULLPROF-2020 software.
\subsection{Magnetization measurements}
The magnetic susceptibility was measured over a temperature range of 5–300 K under applied magnetic fields of 100 Oe, 1000 Oe, and 10 kOe, considering both zero-field-cooled (ZFC) and field-cooled (FC) conditions. Magnetization measurements were conducted at 5 K in a field range of -9 T to 9 T.

\subsection{Transport measurements}
The temperature dependence of the Seebeck coefficient for Ca$_2$CoNbO$_6$ samples in contact with platinum was investigated in air between 350 and 550 K using a custom-built apparatus. The measurements were conducted with a temperature gradient of 30 K across the sample's edges.

Resistance was measured in a temperature range of 300–600\,K in both zero and 5\,T magnetic fields. Measurements were carried out using the standard 4-contact method, with aluminum wires bonded to the sample surface using silver paste (Westbond, USA).

\subsection{ESR measurements}
ESR measurements were carried out using a Bruker ELEXSYS E500-CW spectrometer, fitted with continuous-flow He and N2 cryostats. Measurements were made in the X-band at a frequency of 9.4 GHz, spanning a temperature range of 5–340 K and a magnetic field range of 0–1.4\,T.

\subsection{DFT calculations}
Firstly, we utilized Quantum ESPRESSO (QE) 6.8 software for our DFT calculations \cite{2009quantum, 2017quantum}. We used PBE exchange-correlation functional \cite{PBE}, and  90 Ry energy cutoff was chosen for the basis set. Norm-conserving pseudopotentials from the AFLOW$\pi$ package were used \cite{aflowpi}, except for the Ca pseudopotential, which was obtained from PseudoDojo \cite{pseudo-dojo-norm}.

In our previous work, we investigated a similar double perovskite, Ba$_2$CoNbO$_6$, using the Agapito-Curtarolo-Stefano-Buongiorno-Nardelli (ACBN0) DFT+$U$ approach to accurately capture the localized nature of the $d$-orbitals \cite{ba2conbo6,acbn0,dft+u}. This method allows for the self-consistent calculation of Hubbard $U$ corrections in DFT + $U$ for each crystallographic position.  For the present calculations, we adopted the average values obtained self-consistently for Ba$_2$CoNbO$_6$: $U_{\text{Co}-3d} = 2.25$ eV and $U_{\text{O}-2p} = 5.8$ eV. A $10 \times 10 \times 8$ $k$-point mesh was used for Brillouin zone sampling.

For each configuration, variable cell relaxation was performed, allowing atomic positions to relax until the residual forces were below 10$^{-3}$\,Ha/Bohr (approximately 0.05\,eV/\AA). Additionally, convergence was achieved when the total energy difference between two successive self-consistent cycles was less than 10$^{-4}$\,Ha.

In addition, we provide single-point calculations on the QE relaxed structures using ACBN0 upon PBE xc-functional. For this calculation we utilized FHI-aims program package \cite{FHIaims} - all-electron full potential code with numeric atom centered orbitals as basis set. We chose Mulliken projection functions definition of localized subspaces for Nb-4d, Co-3d and O-2p electronic shells. The resulting U values for each structure can be found in the appendix \ref{appendix} (see Table \ref{U_ACBN0}). It is worth noting that in FHI-aims DFT+U implementation, there is an opportunity to use Petukhov mixing factor \cite{Petukhov} in order to find a balance between the fully localized limit (FLL) and around mean-field (AMF) limit for treatment of the double counting term for the DFT+U approach. However, numerous computational experiments showed that the standard FLL limit for double counting term is the most preferable choice for the bulk structures of strongly correlated materials since it accounts for strong localization for transition metal electrons properly, while the mixing parameter tends to slightly delocalize it. The $15 \times 15 \times 12$ $k$-point mesh and "light" basis set functions with "tight" grids for integration were used. The gaussian broadening function width for Fermi level occupation distribution was set to 0.01 eV. The accuracy settings are the following - $10^{-3}$ eV for the sum of Hamiltonian eigenvalues, $10^{-5}$ eV for the total energy, and $10^{-4}$ for the charge density.

We analyzed two possible distributions of transition metals within the structure, as shown in Fig.~\ref{srtuc} b-c. The first scenario assumes a rock-salt distribution, where each niobium atom has cobalt as its nearest transition metal and vice versa. In the second scenario, alternating layers of cobalt and niobium are arranged.  

\section{Results and discussions}
\subsection{Crystal structure and cation ordering}
\subsubsection{XRD analysis}
XRD analysis confirmed the formation of the double perovskite Ca$_2$CoNbO$_6$. The diffraction pattern was indexed with the space group \( P12_1/c1 \) (No.~14), corresponding to a monoclinic crystal system, consistent with the results reported by Shaheen and Bashir~\cite{Shaheen}. In their work, a 70\%/30\% Co/Nb ordering was assigned to the 2c/2d Wyckoff positions, respectively, based primarily on the (101) reflection (in the \( P2_1/n \) notation), which is sensitive to cation ordering.  

\begin{figure}[t]
\centering
\includegraphics[width=0.7\linewidth]{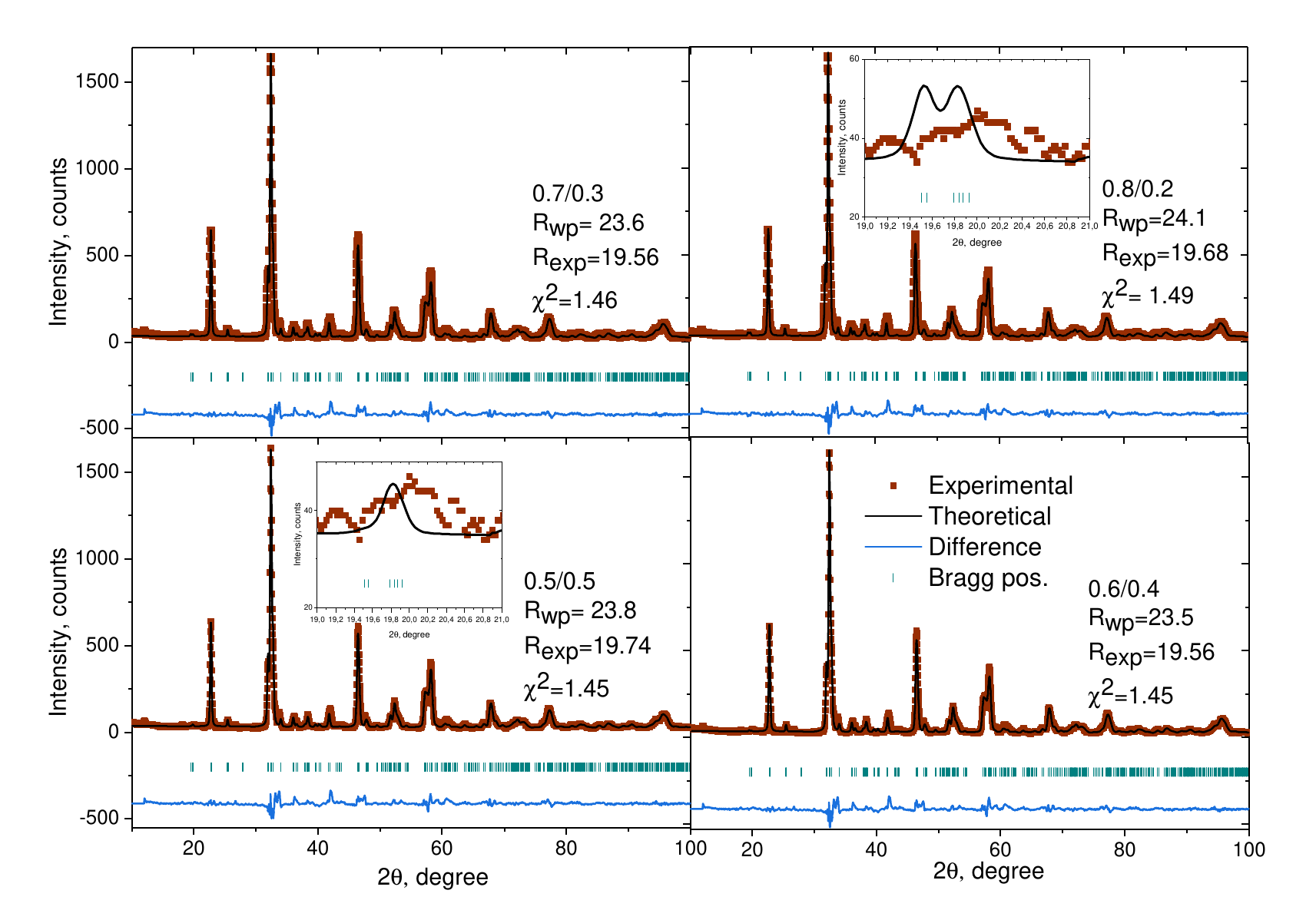}
\caption{Results of the XRD measurements,  Rietveld refinement, the difference between refinements and experimental data, and the theoretical positions of Bragg reflections for the space group \( P12_1/c1 \). The inset shows the  the (101) reflection corresponding to the presence of ordering of cations.}
\label{xrd-fig}
\end{figure}

The results of the refinement are presented in Figure \ref{xrd-fig} and Table \ref{xrd-table}. In our case, we were unable to achieve an accurate fit in this region (see the insets in Figure \ref{xrd-fig}), although the overall refinement fits well. We tested several Co/Nb site-occupation ratios ranging from 0.5/0.5 to 0.8/0.2. The corresponding values \(\chi^2\) varied only marginally between 1.45 and 1.49, with the minimum obtained at 0.5/0.5. Ratios up to 0.65/0.35 produced essentially indistinguishable fits (\(\chi^2\) = 1.45), with differences emerging only beyond the third decimal place. It is also important to note that XRD analysis alone does not provide a definitive basis for determining occupancy factors. For instance, in \cite{La4LiMnO8}, Li and Mn were not distinguishable by XRD, whereas NMR measurements indicated the presence of Li/Mn ordering in the sample. 

\begin{figure}
    \centering
    \includegraphics[width=0.3\linewidth]{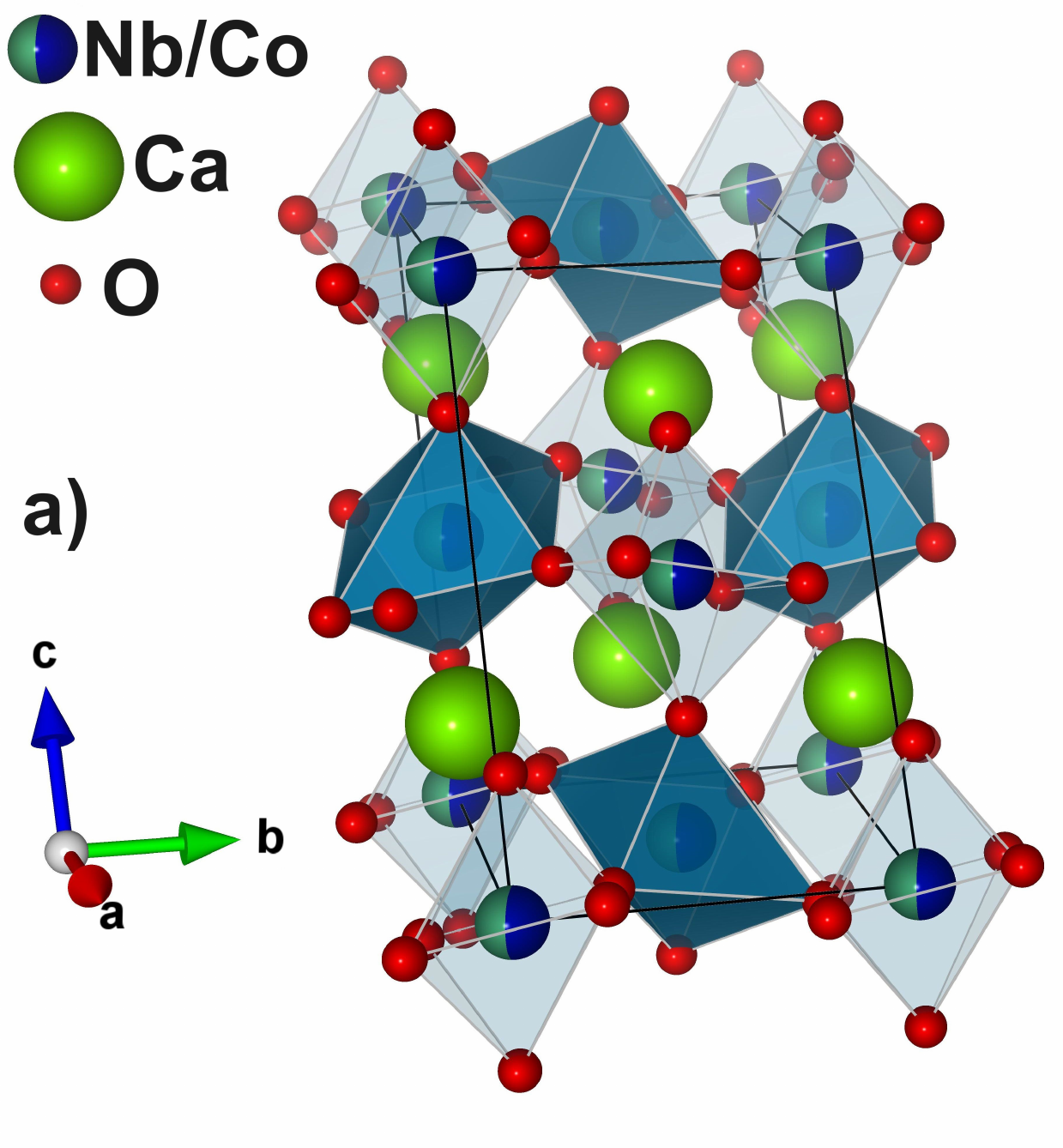}
    \includegraphics[width=0.3\linewidth]{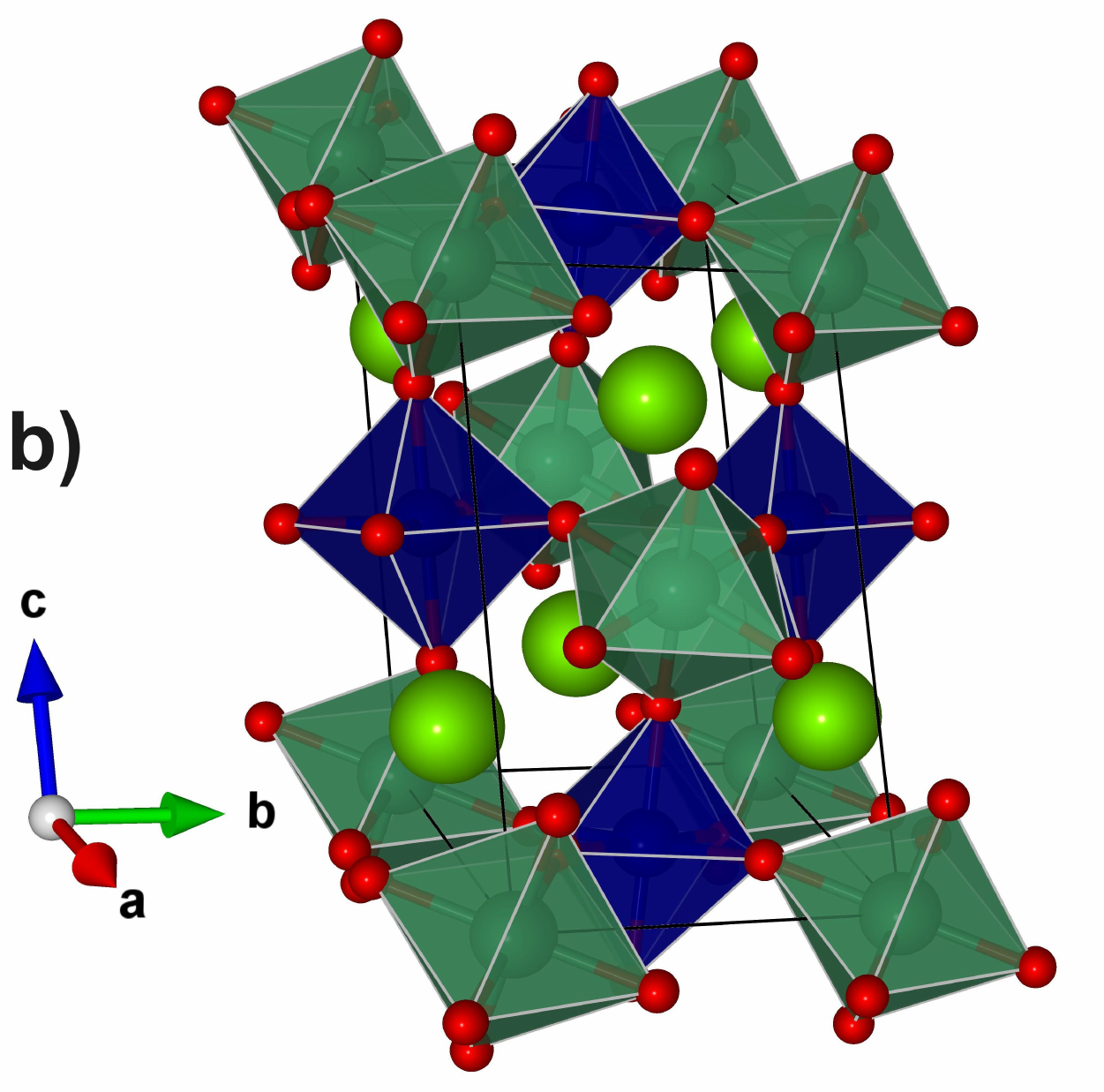}
    \includegraphics[width=0.3\linewidth]{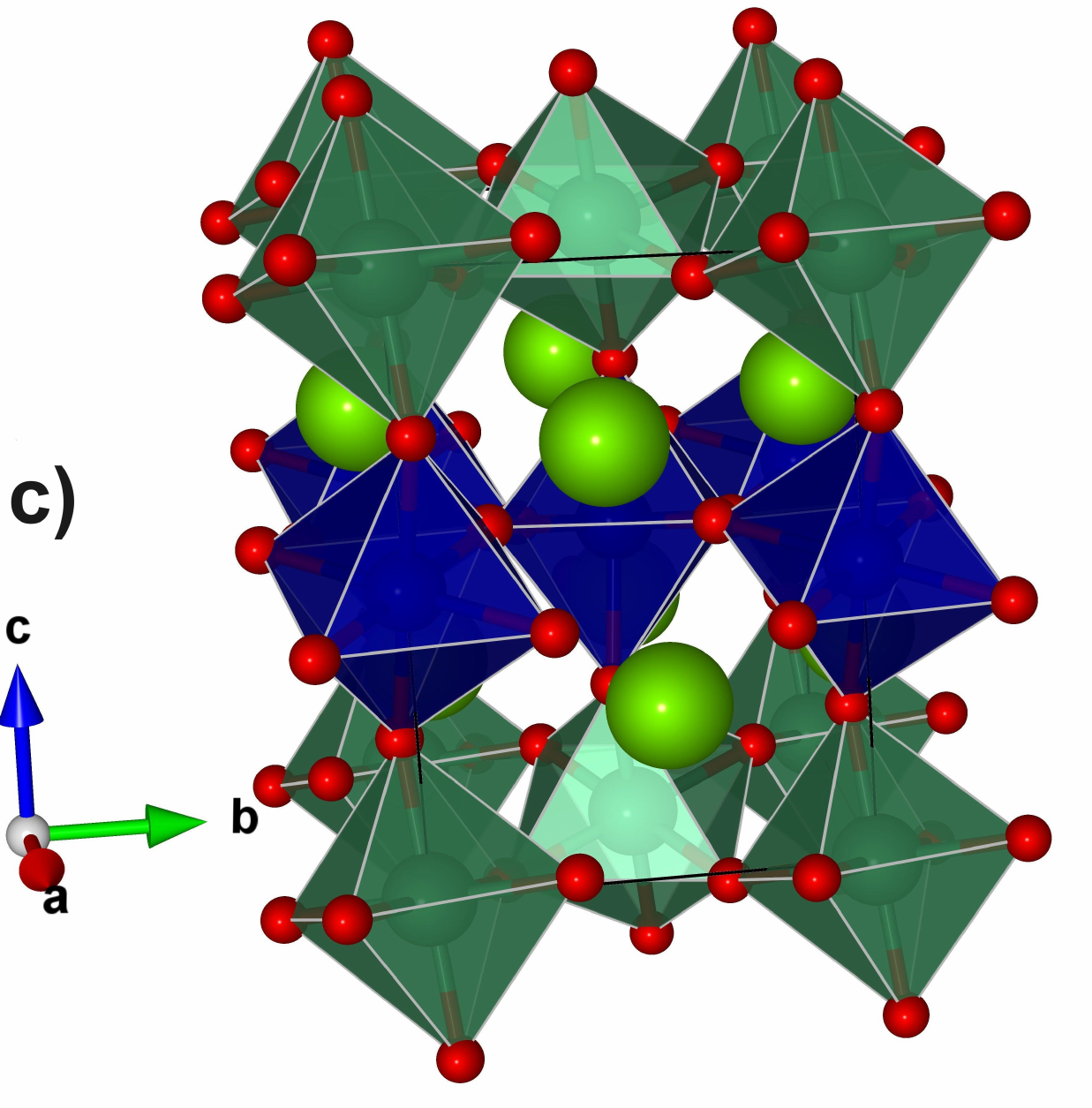}
    \caption{Crystal structure of Ca$_2$CoNbO$_6$. Dark green represents niobium, blue represents cobalt, light green represents calcium, and red represents oxygen. (a) Experimentally determined structure, no ordering of Co/Nb; blue-green polyhedra with varying transparency correspond to the two nonequivalent positions of Co/Nb. (b) DFT-relaxed structure with a rock-salt (uniform) distribution of Co and Nb. (c) DFT-relaxed structure with alternating layers of Co and Nb. Visualization was performed using VESTA software~\cite{vesta}.
   }
    \label{srtuc}
\end{figure}

The crystal structure is illustrated in Figure \ref{srtuc}a, showing the anti-phase tilting along [100]$_P$ and the in-phase tilting along [101]$_P$ of adjacent octahedra. This tilting arises due to the small ionic radius of Ca$^{2+}$ cations, which forces the CoO$_6$ and NbO$_6$ octahedra to tilt in order to optimize the Ca–O bond lengths. As a result, this tilting induces octahedral distortion, leading to two non-equivalent crystallographic positions occupied by Co/Nb. 

\begin{table}[width=0.8\linewidth,cols=6,pos=t]
    \caption{Structural characteristics  of the double perovskite Ca$_2$CoNbO$_6$ and  reliability factors obtained from the Rietveld refinement} \label{xrd-table}
    \begin{tabular*}{\tblwidth}{ccccccc}
    \toprule
        \textbf{Cell parameters} & $a$ (\AA) & $b$ (\AA) & $c$ (\AA) & $V$ (\AA$^3$) & $\beta,\degree$  \\
        \midrule
        & 5.4761(4) & 5.6083(4) & 9.5459(1) & 228.88(4) & 125.095(7) \\
        \midrule
\textbf{Atomic positions} & $x$ & $y$ & $z$  & Wyckoff positions \\
        \midrule
        Ca & 0.243(4) & 0.471(1) & 0.243(2)  & 4e \\
        Nb1/Co1 & 0.5 & 0 & 0.5  & 2a \\
        Nb2/Co2 & 0 & 0 & 0 & 2d \\
        O1 & 0.417(7) & 0.156(4) & 0.123(4)  & 4e \\
        O2 & 0.264(5) & 0.701(2) & 0.031(2)  & 4e \\
        O3 & 0.155(7) & 0.045(4) & 0.246(5)  & 4e \\
        \midrule
        \textbf{Reliability factors} & $R_f$ & $R_p$ & $R_{wp}$ & $R_{exp}$ & $\chi^2$ \\
        \midrule
        & 4.94 & 8.05 & 14.2 & 11.94 & 1.41 \\
        \bottomrule
    \end{tabular*}
\end{table}

\subsubsection{DFT calculations}
To investigate the crystal structure of Ca$_2$CoNbO$_6$, we conducted DFT calculations for two different distributions of Co/Nb: (i) a rock-salt (uniform) distribution of Co and Nb (Figure \ref{srtuc} b), and (ii) alternating layers of Co and Nb (Figure \ref{srtuc} c). In principle, the \textit{ab initio} methods provide information only about thermodynamic stability and cannot explicitly confirm or exclude the presence of metastable distributions in experimental structures \cite{lasrco1/2fe1/2o4}. However, in our case, some insights could be obtained. 

The relaxed structures are presented in Figure~\ref{srtuc}. 
The crystallographic angle $\beta$ becomes equal to $90^\circ$ for the structure with interchanged Co and Nb layers. 
At first glance, it may appear that this structure no longer belongs to the space group $P12_1/c1$. 
However, it still retains the $P12_1/c1$ symmetry; after relaxation, Nb occupies the $2a$ Wyckoff position and Co the $2b$ position (which was not the case prior to relaxation, both in terms of symmetry and atomic positions).  In contrast, for the rock-salt distribution case, Nb occupies the $2a$ position while Co occupies the $2d$ position. The relaxation results clearly show that in-phase tilting is energetically more favorable for elements with different ionic radii.  The structure with interchanged layers is lower in energy, which may indicate that anti-phase tilting between elements of the same ionic radius can be compensated by magnetic ordering.

For further analysis, we assumed that O is in the $2^-$, Ca in the $2^+$, Nb in the $5^+$, and Co in the $3^+$ oxidation states, as in this case, all elements are in their stable oxidation states, and charge neutrality is satisfied. Consequently, Nb$^{5+}$ is non-magnetic due to its $4d^0$ configuration, and all magnetism arises from Co. For the analysis of spin states from QE, we used  L{\"o}wdin magnetic moments, which represent the difference between the number of spin-up and spin-down electrons projected onto an atom using  L{\"o}wdin partitioning. For FHI-aims results, we use Mulliken magnetic moments. In the $3^+$ oxidation state, Co has a $3d^6$ configuration and can adopt a high-spin state with $S=2$, an intermediate-spin state with $S=1$, or a low-spin state with $S=0$. The obtained magnetic L{\"o}wdin  moments are presented in Table \ref{dft-table} and Mulliken moments in Table \ref{acbn0_tab}. Magnetic moments of approximately 3 could, in principle, correspond to Co in the intermediate spin state or in the high spin state. However, the total magnetic moment of 8 per unit cell (per 2 Co atoms) indicates that Co is in the high-spin state. The discrepancy between L{\"o}wdin magnetic moment and the total magnetic moment per unit cell  arises from hybridization of Co-$3d$ orbital with O-$2p$  \cite{lasrco1/2fe1/2o4}. 

To obtain the FM-ordered state for the layered structure in QE calculations, we constrained the total magnetization within the unit cell. Even after relaxing the geometry under this constraint, adjusting the initial magnetic moments, and tuning the mixing parameters, we were unable to achieve convergence of the self-consistent field (SCF) cycle for the FM configuration without enforcing magnetization constraints. At the same time, we were unable to obtain Co in the intermediate spin state (IS) in QE calculations despite adjusting mixing parameters and initial magnetic moments. In FHI-aims calculations, the intermediate spin state for the FM configuration in the layered structure SCF cycle was converged without any additional constraints, while the high-spin state required a total moment constraint. For this ordering, we see that IS state lies lower than HS state by $\approx 0.1$ eV/u.c., at the same time, it is the only structure that is metallic within our approximations (see Fig. \ref{ACBN0_FM_pDOS}). The coexistence of metallic conductivity and a small energy difference between IS/HS indicates the itinerant character of magnetism in the layered FM structure.

For both considered orderings, the nonmagnetic configuration corresponding to Co in the low-spin state is highly energetically unfavorable. Therefore, according to the DFT calculations, Co is in the high-spin state in the double perovskite Ca$_2$CoNbO$_6$.

The energy difference between the FM and AFM configurations is nearly 60 times larger for the structure in which the Co and Nb layers are interchanged than for the structure with rock-salt-type distributed Co and Nb (see Table \ref{dft-table}). These results indicate strong AFM coupling in the layered configuration and weak AFM coupling in the rock-salt distributed configuration. The weak AFM coupling likely corresponds to a paramagnetic regime observed experimentally. In FHI-aims calculations, we see that the AFM/FM energy difference is of the same order. This may be the effect of the absence of relaxation in our FHI-aims calculations, therefore it should not be treated as a quantitative measure but only as a qualitative result of the distribution of different magnetic phases.

\begin{table}[width=0.65\linewidth,cols=7,pos=t]\caption{Total energy differences, crystallographic angle $\beta,$ and L{\"o}wdin magnetic moments obtained via DFT+U in QE.  Energy difference is presented with respect to the structure with rock-salt distribution of Co and Nb with antiferromagnetically ordered Co.}
\label{dft-table}
\begin{tabular*}{\tblwidth}{lcccccc}
\toprule
& \multicolumn{3}{c}{\textbf{rock-salt}} & \multicolumn{3}{c}{\textbf{layered}} \\
\midrule
\textbf{Magnetic ordering}              & AFM     & FM      & NM      & AFM             & FM             & NM            \\
\midrule
$\Delta E,$ eV                 & 0       & 0.007   & 0.35    & -0.033          & 0.36           & 0.82          \\
$\beta,$\degree & 125.67  & 125.67  & 125.46  & 90.00           & 90.02          & 90            \\
$\mu_{Co}, \mu_B$                     & 3.1     & 3.1     & 0       & 3               & 3              & 0  \\    
\bottomrule
\end{tabular*}
\end{table}

\begin{table}[width=0.7\linewidth,cols=8,pos=t]    \caption{Total energy differences and Mulliken magnetic moments per Co atom obtained via DFT+U and ACBN0 in FHI-aims. Energy difference is presented with respect to the structure with roack-salt distribution of Co and Nb with antiferromagnetically ordered Co.}
    \label{acbn0_tab}
    \centering
    \begin{tabular}{lccccccc}
    \toprule
 & \multicolumn{3}{c}{\textbf{rock-salt}} &  \multicolumn{4}{c}{\textbf{layered}} \\ 
 \midrule
 \textbf{Magnetic ordering} & AFM & FM & NM & AFM & FM (IS) & FM (HS) & NM \\ 
 \midrule
$\Delta E$, eV  & 0 & 0.74 & 2.21 & -0.045 &0.65 & 0.74 &	2.33  \\
$\mu_{Co}$, $\mu_B$   & 3.03 &	3.05 & 0 & 2.93 & 2.55 & 2.95 &	0 \\ 
\bottomrule
\end{tabular}
\end{table}

The calculated densities of states for the AFM configuration from QE calculations are presented in Figure \ref{DOS}, from FHI-aims in Figure \ref{ACBN0_pDOS}. The calculated band structures are presented in the appendix \ref{appendix} (see Figures \ref{ACBN0_bands},\ref{ACBN0_bands_FM_layers}) The valence band is mainly formed by hybridized states of oxygen and cobalt atoms in both structures. States associated with Nb atoms are almost absent in the valence band, indicating the ionic nature of the Nb–O bond. This observation suggests that electrical conductivity primarily occurs through Co–O–Co pathways. In the layered structure, this results in insulating behavior of the Nb planes, whereas in a rock-salt distributed Co/Nb structure, the conductivity regime would likely occur via hopping between Co atoms. 
\begin{figure}
    \centering
    \includegraphics[width=0.9\linewidth]{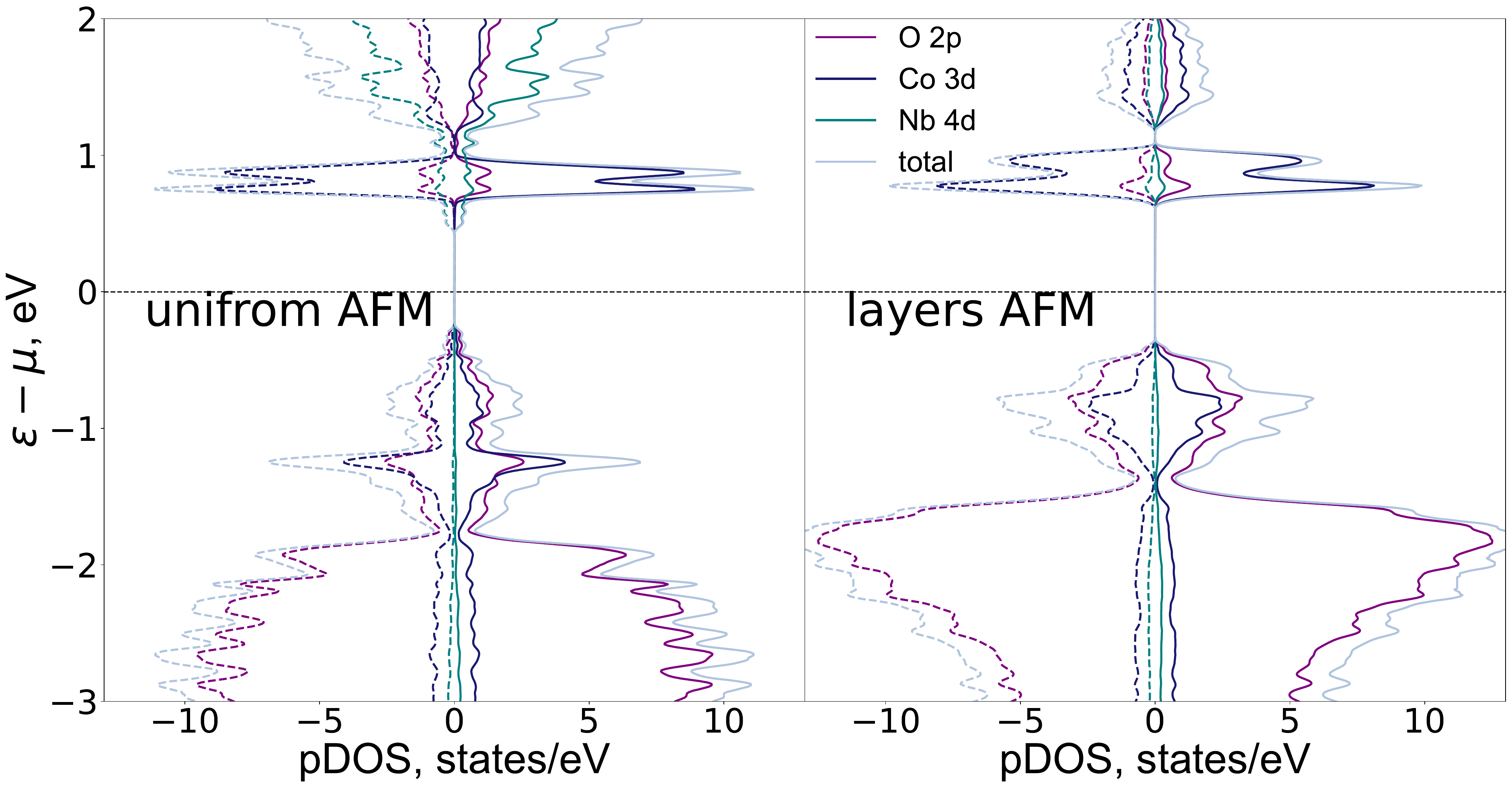}
    \caption{Projected density of states for Ca$_{2}$CoNbO$_{6}$ antiferromagnetic state for rock-salt (uniform) (left) and layered (right) distribution of Co/Nb ions }
    \label{DOS}
\end{figure}

\subsection{Macroscopic magnetic properties}
\begin{figure}
    \centering
    \includegraphics[width=0.5\linewidth]{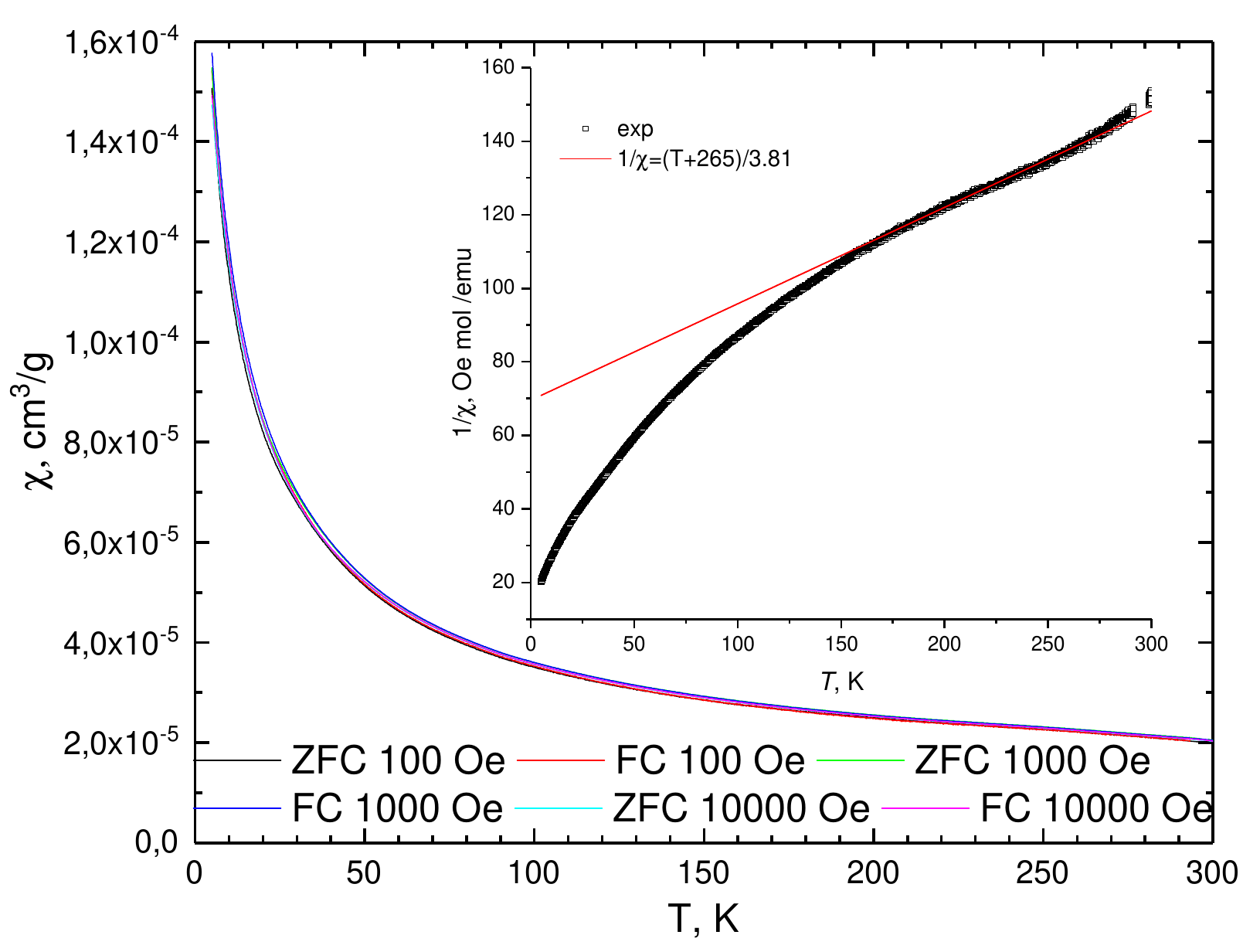}
    \caption{Magnetic susceptibility measured in various regimes. The inset shows the inverse susceptibility along with the Curie-Weiss fit}
    \label{chi-fig}
\end{figure}

 The results of the susceptibility measurements are shown in Figure \ref{chi-fig}. No evidence of magnetic transitions was detected across the measured temperature range. The absence of magnetic ordering indicates a rock-salt distribution of Co and Nb ions, analogous to the double perovskite Sr$_2$CoNbO$_6$ \cite{sr2conbo6}, where X-ray diffraction (XRD) confirmed homogeneous cation ordering. In contrast, the presence of Co--O--Co linkages would facilitate superexchange interactions capable of inducing magnetic ordering, as reported for other double perovskites such as Ba$_2$CoNbO$_6$ \cite{ba2conbo6}.

In our case, however, the magnetic susceptibility deviates from ideal Curie-Weiss paramagnetic behavior below approximately 150~K, indicating the presence of magnetic correlations that are not sufficiently strong to establish long-range magnetic order.

At high temperatures, the system behaves paramagnetically, and the susceptibility follows the Curie-Weiss law:
\begin{equation}
    \chi=\frac{C}{T-\theta_{CW}}.
\end{equation}
Here, $C$ is the Curie-Weiss constant and $\theta_{\mathrm{CW}}$ is the Curie-Weiss temperature. The value of the Curie-Weiss constant obtained from the fitting results is $C = 3.81$ emu K/Oe mol. The Curie-Weiss constant is related to the magnetic moment by the following formula: $\sqrt{\frac{3k_B C}{N_A}}$, where $k_B$ is the Boltzmann constant and $N_A$ is the Avogadro constant. Using this relation, the experimental magnetic moment is calculated as $\mu_{\mathrm{exp}} = 5.52 \, \mu_B$. This value is consistent with the typical magnetic moment observed for Co in a high-spin state within an octahedral ligand field~\cite{sus-tutorial}.

The theoretical magnetic moment of a system of localized spins can be estimated using the formula $\mu_{\text{theor}} = g \sqrt{\sum_{i} N_i \cdot S_i \cdot (S_i + 1)}$, where the sum is applied to the magnetic ions, $N_i$ represents the number of magnetic ions per unit cell and $S_i$ is the spin. For Co$^{3+}$ in the high-spin state, using a $g$-factor of 2.24 obtained from EPR measurements (see below), the theoretical magnetic moment is 5.48 $\mu_B$.

\begin{figure}
    \centering
    \includegraphics[width=0.5\linewidth]{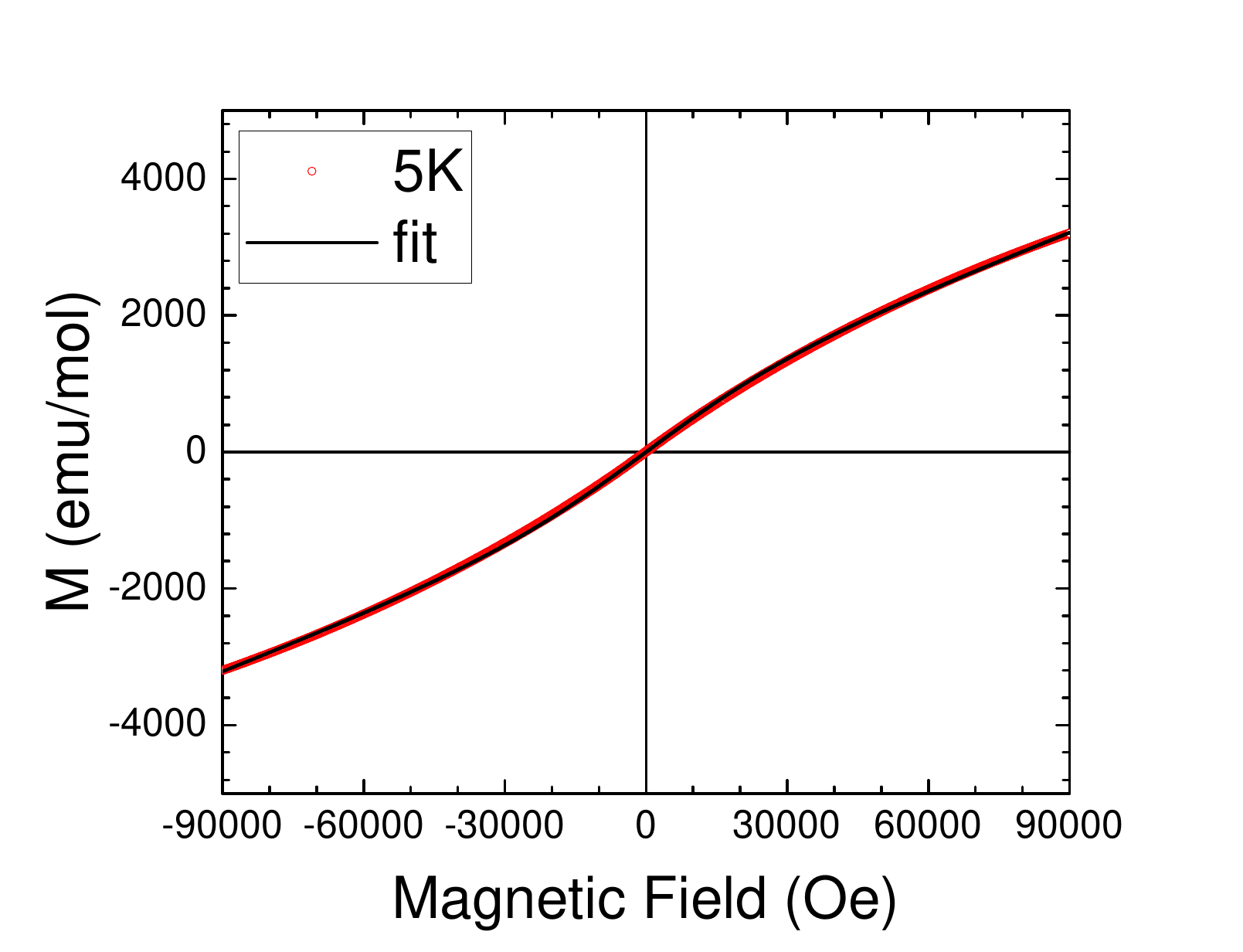}
    \caption{Magnetization versus magnetic field at 5\,K in Ca$_{2}$CoNbO$_{6}$}
    \label{M-H}
\end{figure}

The dependence of magnetization on the external magnetic field is shown in Figure~\ref{M-H}. As can be seen, the magnetization does not saturate within the measured magnetic-field range. The effective magnetic moment per formula unit at the maximum applied field of 9~T is equal to $0.47\,\mu_\mathrm{B}$, which is significantly smaller than the expected saturation magnetization of $4\,\mu_\mathrm{B}$ for high-spin Co$^{3+}$ ($S=2$).

To describe the magnetization, we use a model consisting of two contributions: a linear term, $\chi H$, characteristic of a system with dominant antiferromagnetic correlations, and a superimposed Brillouin-like contribution describing weakly interacting paramagnetic spins expected to saturate in high magnetic fields:
\begin{equation}
    M = NgS\mu_\mathrm{B} B_S(x) + \chi H,
\end{equation}
where
\begin{equation}
    x = \frac{gS\mu_\mathrm{B}H}{k_\mathrm{B}T},
\end{equation}
and
\begin{equation}
    B_S(x)=\frac{2S+1}{2S}\coth\left(\frac{2S+1}{2S}x\right)
    -\frac{1}{2S}\coth\left(\frac{x}{2S}\right)
\end{equation}
is the Brillouin function. Here, $N$ is the number of magnetic ions per mole, $g$ is the spectroscopic splitting factor, and $\chi H$ represents the linear susceptibility contribution.

The best agreement with the experimental data was obtained for $S=3/2$, corresponding to high-spin Co$^{2+}$. The presence of Co$^{2+}$ ions is independently confirmed by the EPR spectra discussed below. The concentration of the paramagnetic centers obtained from the fit is approximately 0.025 per formula unit. We also attempted to fit the data assuming $S=2$, corresponding to high-spin Co$^{3+}$ previously established for Ca$_2$CoNbO$_6$ from susceptibility measurements; however, the quality of the fit was significantly worse.

The low-temperature paramagnetic contribution indicates the presence of magnetically weakly interacting spins, which are associated with locally ordered regions where the nearest-neighbor B-site cations surrounding Co are predominantly Nb. Based on the Rietveld refinement results, the probability of forming such regions is approximately $0.6^6 \approx 0.047$, which is close to the concentration of paramagnetic Co$^{2+}$ centers obtained from the magnetization fit. This suggests that Co$^{2+}$ ions preferentially occupy regions with locally ordered Co/Nb distribution.

\subsection{Electrical transport and thermoelectric properties}
\begin{figure}
    \centering
    \includegraphics[width=0.47\linewidth]{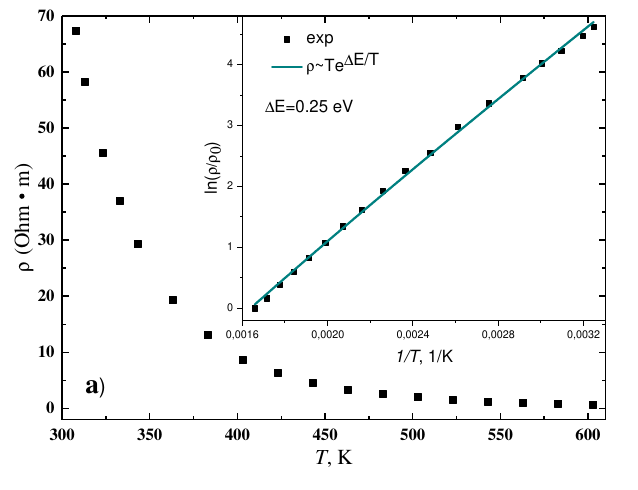}
    \includegraphics[width=0.47\linewidth]{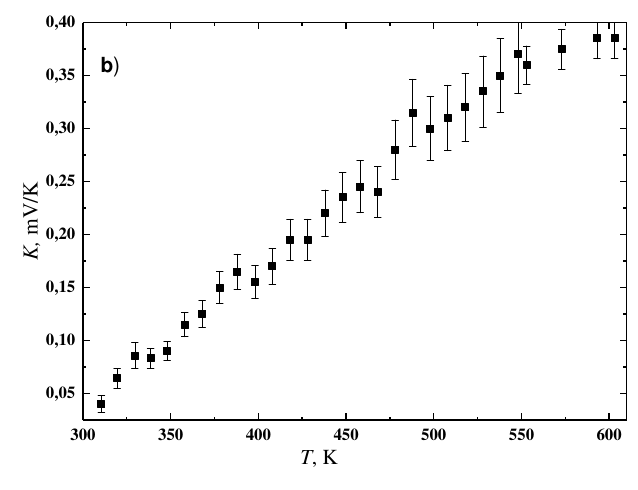}
    \caption{Results of transport measurements: (a) Resistivity and (b) Seebeck coefficient. The inset in (a) shows the resistivity fit using the small polaron hopping model}
    \label{rho-seebeck}
\end{figure}

The results of the transport measurements are presented in Figure~ \ref{rho-seebeck}. Electrical conductivity exhibits thermally activated behavior, increasing with temperature in the range 300–600 K. The conductivity is described best by the small-polaron hopping model:
\begin{equation}
    \sigma=\text{const}+\frac{A}{T}e^{-\frac{\Delta E}{k_{B}T}},
\end{equation}
where $A$ is a constant $\Delta E$ is the band gap.  Fitting the experimental data yields 
$\Delta E=0.25$ eV, which is very similar to the value reported  for Ba$_2$CoNbO$_6$ double perovskite~\cite{ba2conbo6}. This result is consistent with the DFT calculations, which indicate localized electronic states associated with Co–O hybridization.

The Seebeck coefficient increases monotonically with temperature, reaching 400 $\mu$V/K at 600 K. The positive sign of the Seebeck coefficient indicates hole-type carriers and supports the hopping conduction mechanism.
 
\subsection{ESR}

\begin{figure}
    \centering
    \includegraphics[width=0.47\linewidth]{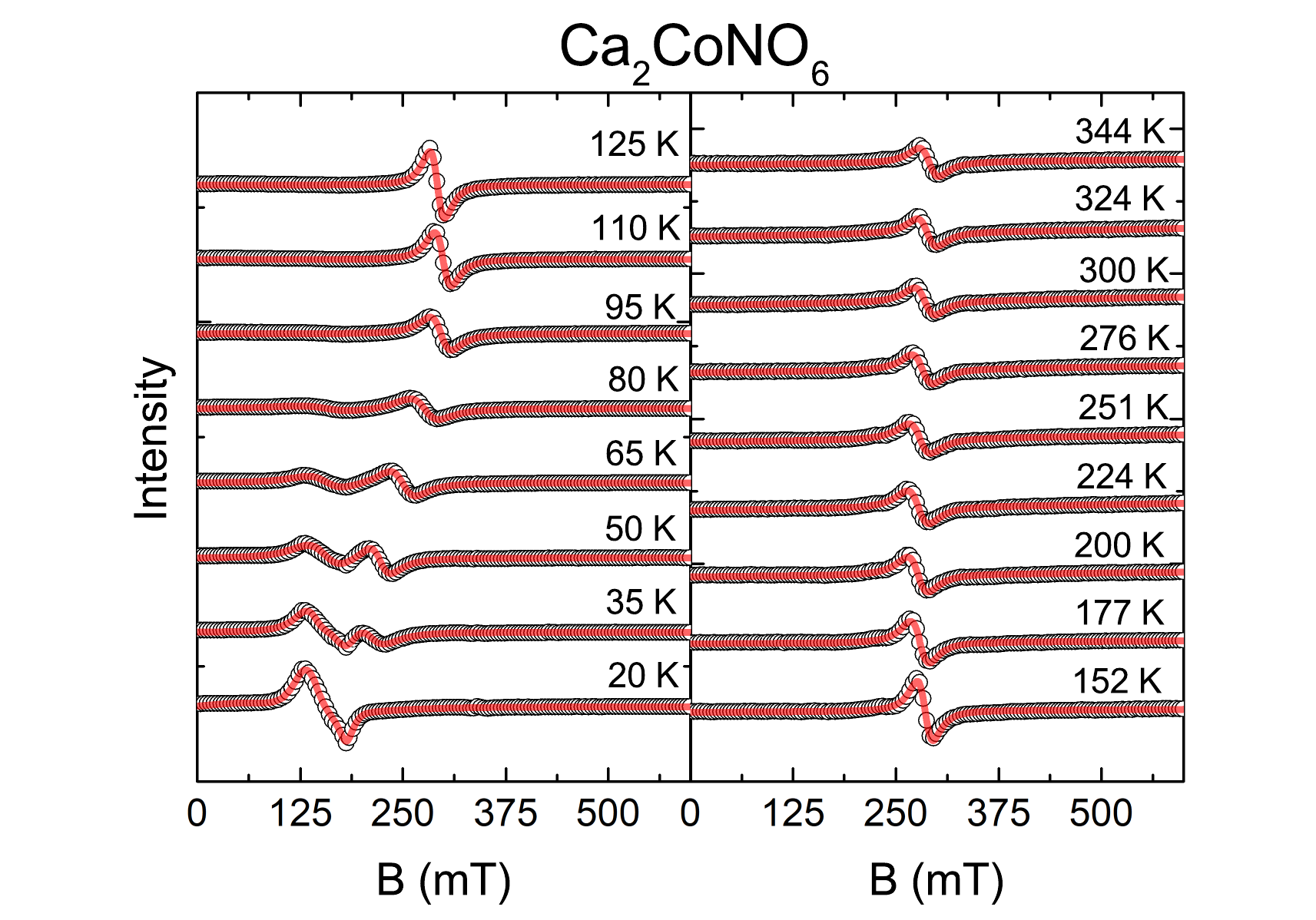}
    \includegraphics[width=0.33\linewidth, angle=90]{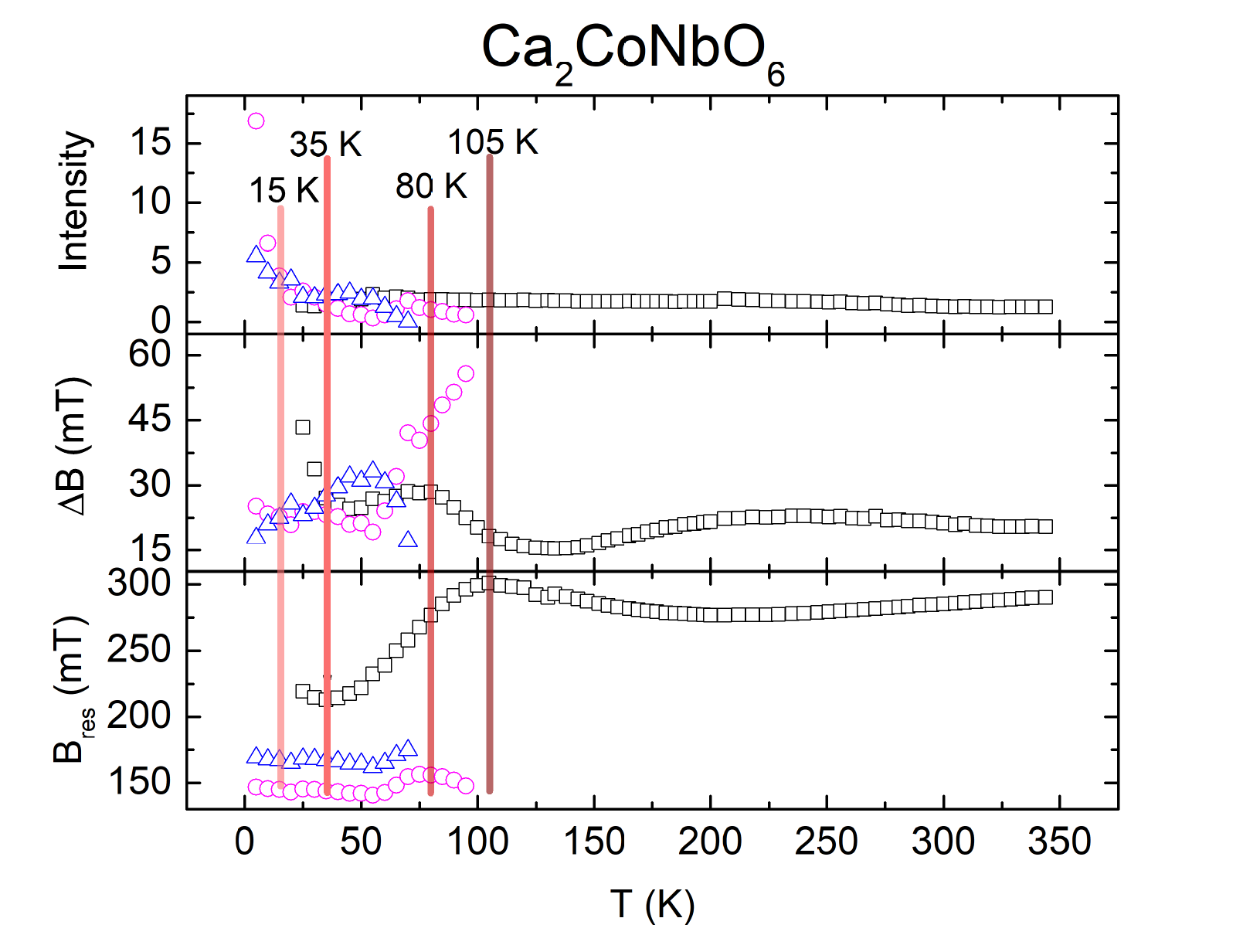}
    \caption{Temperature dependencies of ESR spectra in Ca$_{2}$CoNbO$_{6}$ (left); temperature dependencies of intensity, linewidth, resonance field of ESR spectra in Ca$_{2}$CoNbO$_{6}$ (right)}
    \label{ESRline}
\end{figure}

\begin{figure}
    \centering
    \includegraphics[width=0.7\linewidth]{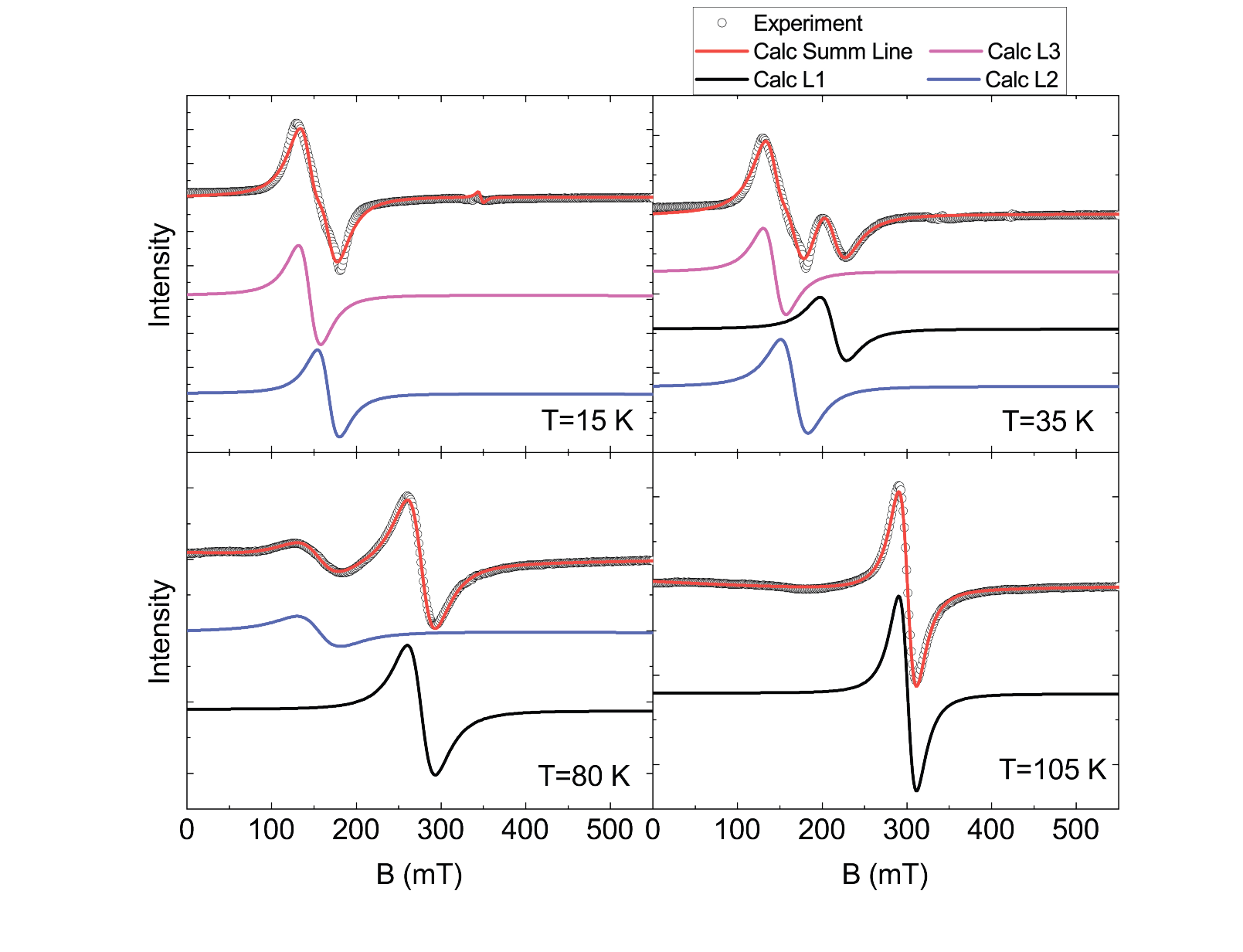}
    \caption{Decomposition of the ESR spectrum into several lines in Ca$_2$CoNbO$_6$ at temperatures of 15, 35, 80, and 105\,K. The symbols represent the experimental data, and the red solid lines represent the sum of the approximation lines, which consist of black, blue, and violet lines depending on the temperature range.}
    \label{ESRseparate}
\end{figure}

The view of the ESR spectra is shown in Figure~\ref{ESRline}. As can be seen from the Figure~\ref{ESRline}, one line is recorded in the ESR spectrum in the temperature range of 344 to 105\,K. As the temperature decreases, the line begins to split and shifts towards low magnetic fields. The ESR spectrum was approximated using two lines in the range of 105 to 85\,K. The ESR spectrum consists of three lines in the range of 85 K to 25\,K. Below 25\,K, only two lines are necessary to describe the shape of the experimental spectrum (see Figure~\ref{ESRseparate}).

The shape of all ESR lines was approximated by the expression:
\begin{equation}
    	{\frac{dP}{dB} = \frac{d}{dB} \Biggl( \frac{\Delta B + \alpha (B - B_\mathrm{res})}{(B - B_\mathrm{res})^2 + \Delta B^2} + \frac{\Delta B + \alpha (B + B_\mathrm{res})}{(B + B_\mathrm{res})^2 + \Delta B^2} \Biggr)}, \end{equation} 
where $B_\mathrm{res}$ is the position of the resonance line, $\Delta B$ is the linewidth, and $\alpha$ is the asymmetry parameter~\cite{Bhat}. Linewidth, intensity, and $B_\mathrm{res}$ inferred from the fitting procedure are shown in Figure~\ref{ESRline}b. The color of the symbols corresponds to the color of the lines in Figure~\ref{ESRseparate}.

Let us consider the temperature behavior of each line individually. The resonance magnetic field of the black line remains practically unchanged as the temperature decreases from 340 to 100\,K. The effective $g$-factor calculated from the relation $h\nu = g_{\mathrm{eff}}\mu_B B_{\mathrm{res}}$ is 2.24. The typical value of the $g$-factor for Co$^{3+}$ in an octahedral environment is close to 2. In our case, however, the octahedra are distorted, which likely affects the $g$-factor value. Upon further lowering the temperature, the resonance magnetic field decreases and the effective $g$-factor increases to approximately 3.1. At even lower temperatures, the line is no longer detected. This behavior indicates a bottleneck regime, in which two ESR centers are coupled but exhibit different temperature dependencies of the relaxation rates. As a result, the observed ESR signal is dominated by one center or the other depending on temperature. In our case, these centers are Co$^{3+}$ and Co$^{2+}$ ions coupled via oxygen. The presence of divalent cobalt is probably related to oxygen non-stoichiometry in the sample. Therefore, the transition observed between 105\,K and 35\,K likely reflects a change in the dominant ESR signal from Co$^{3+}$ to Co$^{2+}.$ The occupation of the spin states with different magnetic moments projections follows the Boltzmann distribution. As the temperature decreases, the population of the lowest-energy states becomes dominant. For Co$^{3+}$ ions, the lowest-energy state corresponds to a zero projection of the magnetic moment along the direction of the applied magnetic field. Consequently, at low temperatures the ESR signal associated with Co$^{3+}$ becomes strongly suppressed and eventually disappears.

The remaining two lines, shown in crimson and blue, correspond to Co$^{2+}$. It is known that the spins of Co$^{2+}$ ions relax rapidly in the lattice and are observable only within a narrow temperature range below 100\,K. The resonance magnetic field values remain practically unchanged, and the effective $g$-factors are 3.95 and 4.57, respectively. These values are typical for Co$^{2+}$ ions occupying different octahedral environments of oxygen ions.  

The presence of two distinct signals associated with Co$^{2+}$ indicates partial ordering in the Co/Nb sublattice. In both a fully ordered and a fully disordered structure, only a single signal would be observed, since all Co atoms would occupy equivalent environments.

\section{Conclusion}
The double perovskite Ca$_{2}$CoNbO$_{6}$ with high Seebeck coefficient of 0.4 mV/K
at 600 K was synthesized using the pyrolysis method of nitrate-organic mixtures of the corresponding components. X-ray diffraction analysis confirmed the formation of double perovskite Ca$_{2}$CoNbO$_{6}$, with the space group P12$_1$/c1 (No. 14). Magnetic susceptibility deviates from paramagnetic behavior below approximately 150 K, indicating the presence of correlations, though not strong enough to induce ordering. We conducted DFT calculations for two different distributions of Co/Nb: (i) a rock-salt distribution of Co and Nb, and (ii) alternating layers of Co and Nb. For both considered orderings, the nonmagnetic configuration corresponding to Co in the low-spin state is highly energetically unfavorable. Therefore, according to the DFT calculations, Co is in the high-spin state in the double perovskite Ca$_{2}$CoNbO$_{6}$ and a strong AFM coupling in the structure where layers of Co and Nb ions are interchanged and a weak AFM coupling in the structure where Co and Nb are rock-salt distributed. The absence of magnetic anomalies in the susceptibility measurements suggests a structure with rock-salt ordered Co and Nb. The experimental magnetic moment is calculated as $\mu_{exp}$ = 5.52 $\mu_\mathrm{B}$. For Co$^{3+}$ in the high-spin state with a g-factor
of 2.24, the theoretical magnetic moment is 5.48 $\mu_\mathrm{B}$.

\appendix

\section{ACBN0 electronic structure data}\label{appendix}

\begin{figure}[h]
    \centering \includegraphics[width=\linewidth]{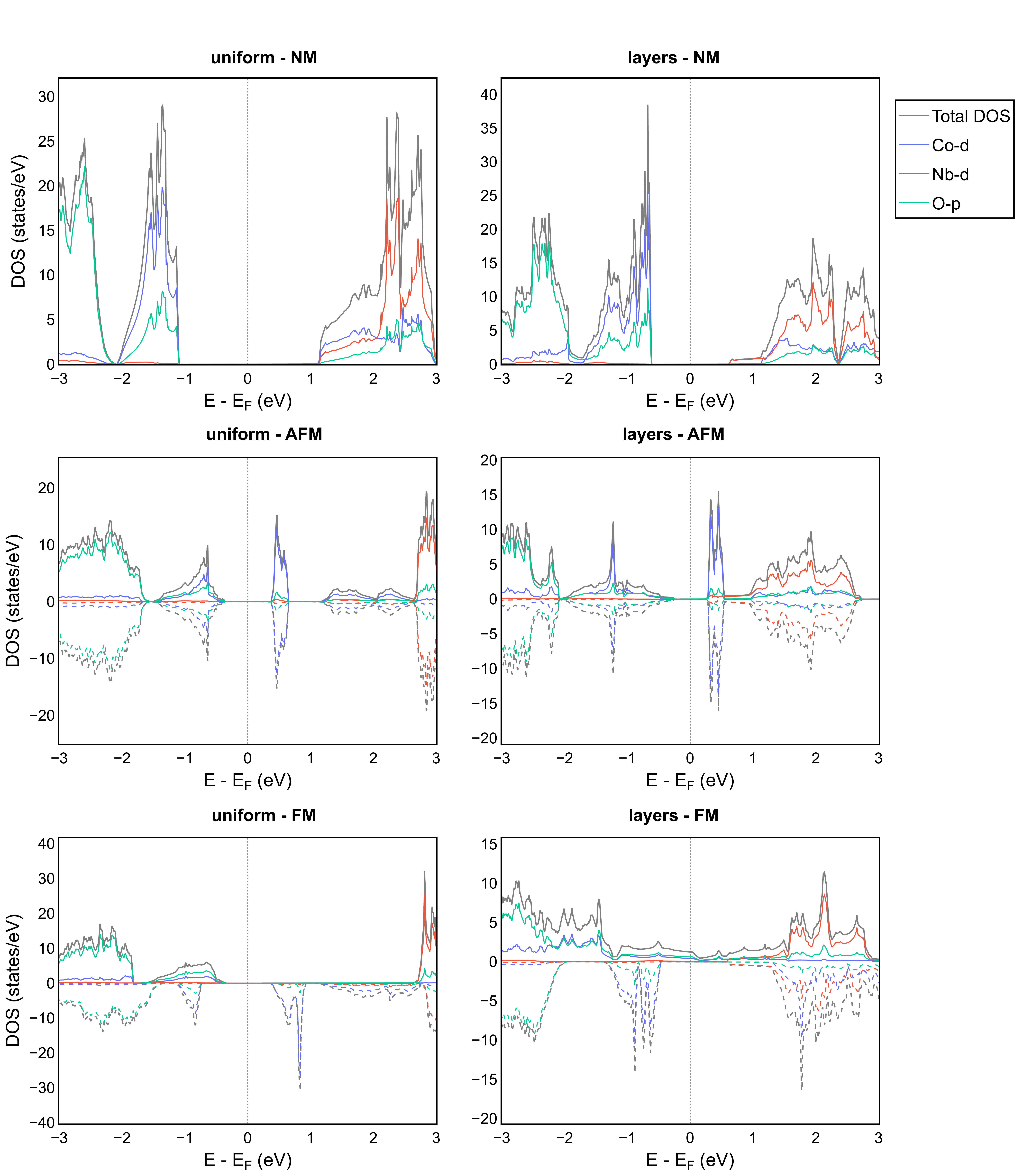}
    \caption{Projected density of states for rock-salt (uniform) (left) and layered (right) Co/Nb distribution and different magnetic orderings calculated with ACBN0@PBE in FHI-aims. For layered FM ordering IS (S=1) state is presented.}
    \label{ACBN0_pDOS}
\end{figure}

\begin{figure}[h]
    \centering  \includegraphics[width=\linewidth]{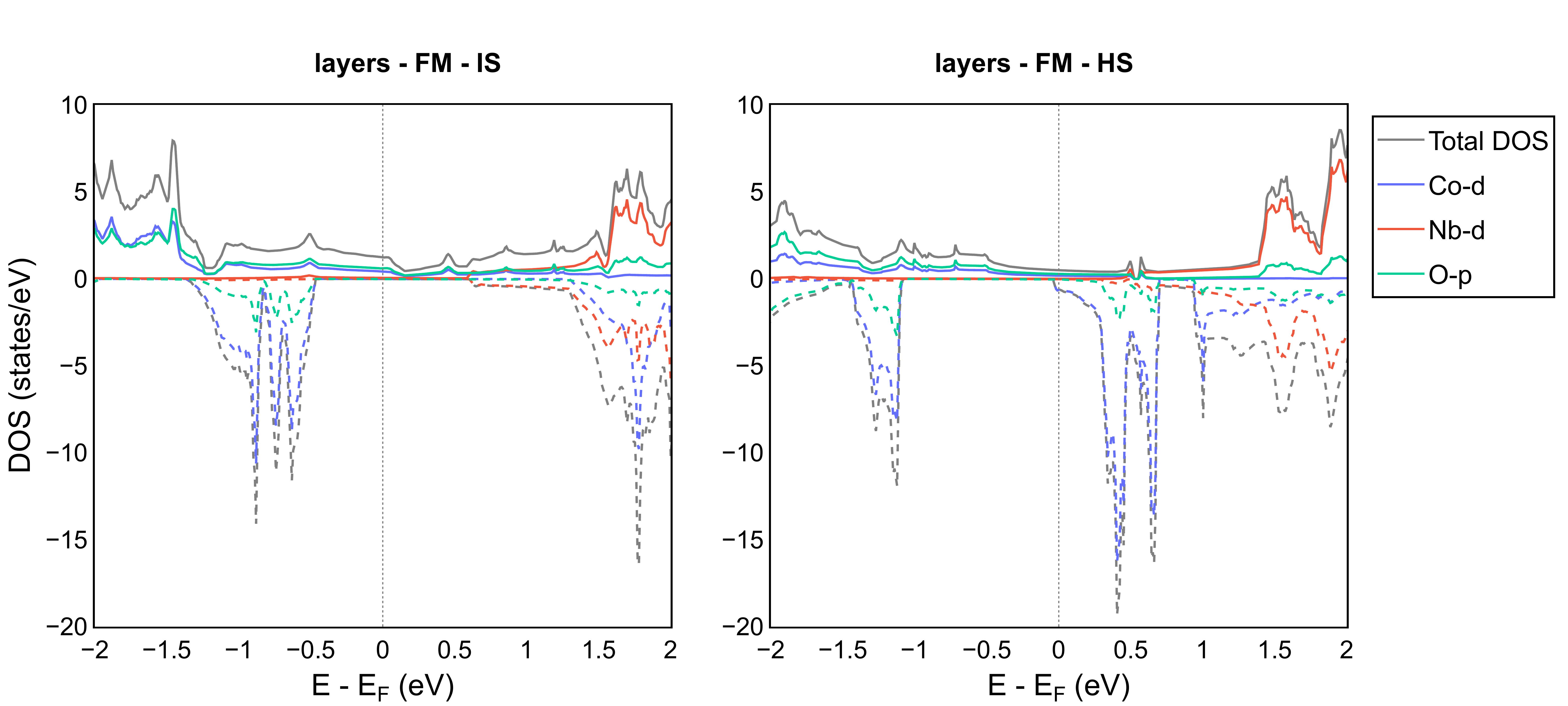}
    \caption{Projected density of states for layered Co/Nb distribution and FM magnetic ordering with different spin state: IS (S=1) (left); HS (S=2) (right)}
    \label{ACBN0_FM_pDOS}
\end{figure}

\begin{figure}[h]
    \centering    \includegraphics[width=\linewidth]{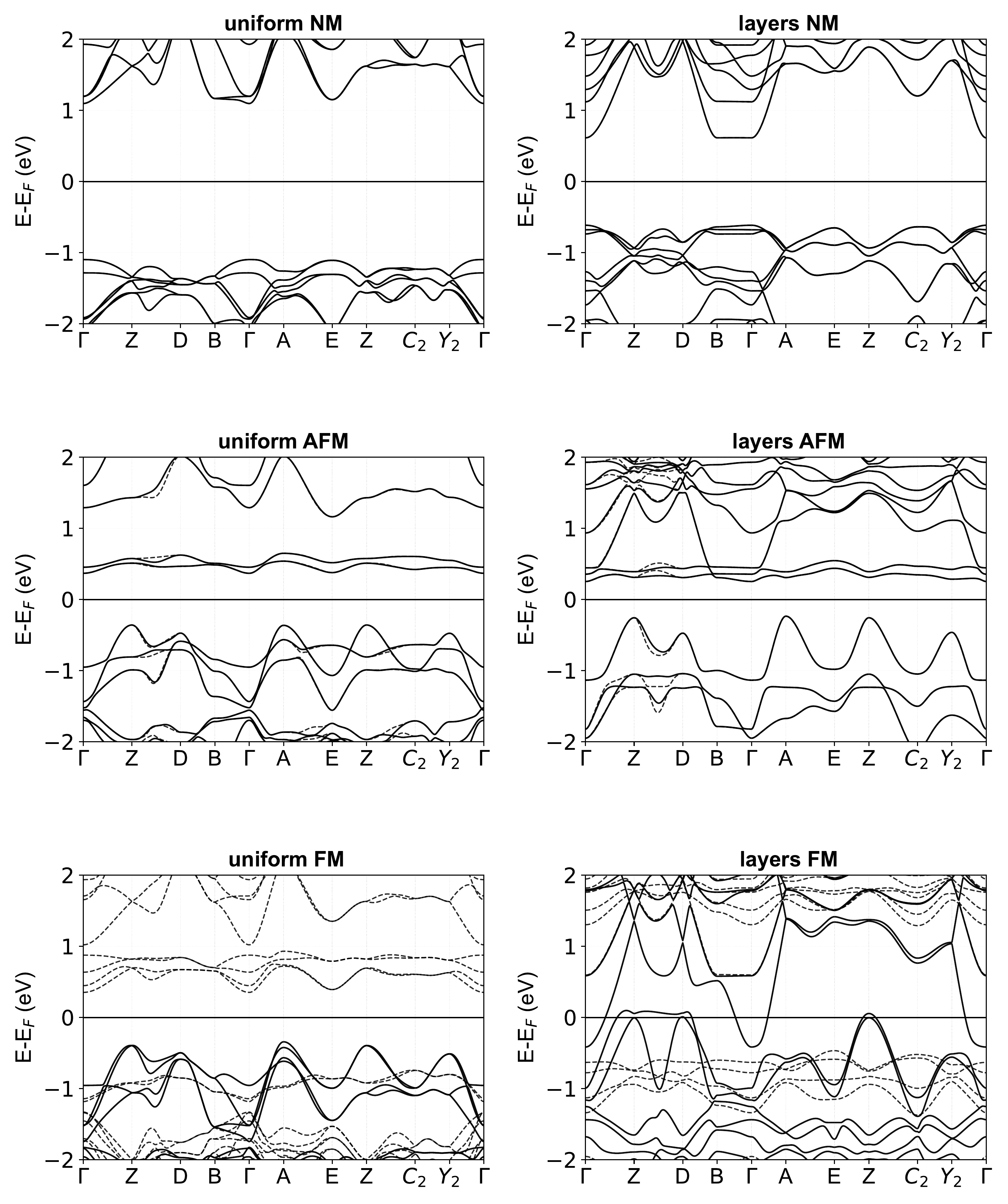}
    \caption{Band structures for rock-salt (uniform) (left) and layered (right) Co/Nb distribution and different magnetic orderings calculated with ACBN0@PBE in FHI-aims. For layered FM ordering IS (S=1) state is presented. Bold lines correspond to the spin-up channel, dashed lines to the spin-down channel.}
    \label{ACBN0_bands}
\end{figure}

\begin{figure}[h]
    \centering    \includegraphics[width=\linewidth]{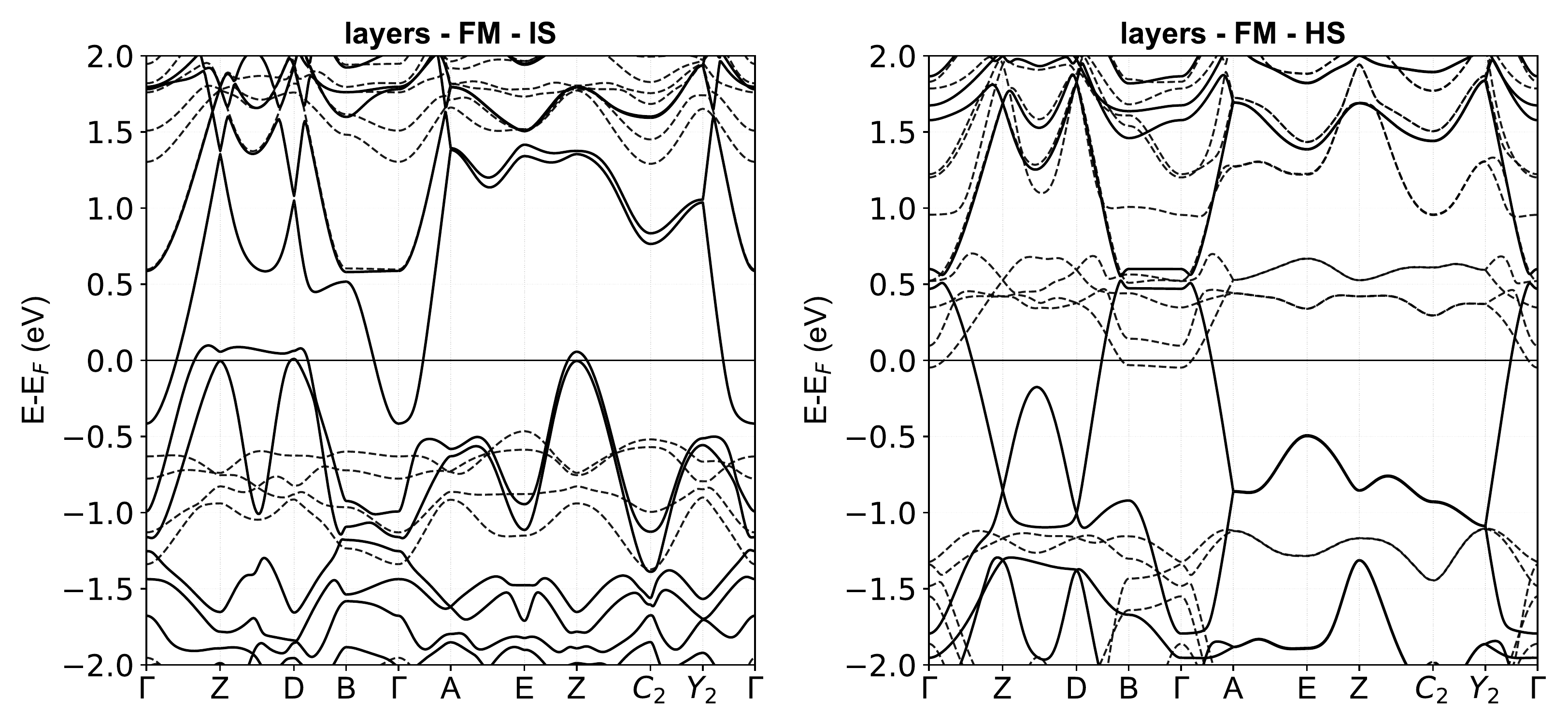}
    \caption{Band structures for layered Co/Nb distribution and FM magnetically ordered structure with IS (S=1) (left) and HS (S=2) (right) spin states calculated with ACBN0@PBE in FHI-aims. Bold lines correspond to the spin-up channel, dashed lines to the spin-down channel.}
    \label{ACBN0_bands_FM_layers}
\end{figure}

The electronic structures of both rock-salt and layered distribution of Co/Nb ions calculated in FHI-aims with DFT+U and ACBN0 are presented in Figures \ref{ACBN0_pDOS}-\ref{ACBN0_bands_FM_layers}.

Figures \ref{ACBN0_pDOS}, \ref{ACBN0_FM_pDOS} present the projected density of states (pDOS) for Ca$_2$CoNbO$_6$ calculated using the ACBN0 self-consistent DFT+U scheme within the FHI-aims all-electron framework for all magnetic configurations and Co/Nb ion distribution: low-spin (NM), antiferromagnetic (AFM), and ferromagnetic (FM). For the layered FM configuration, both intermediate-spin (IS, S = 1) and high-spin (HS, S = 2) of Co$^{3+}$ states were additionally examined (see Figs. \ref{ACBN0_FM_pDOS},\ref{ACBN0_bands_FM_layers}).

In the NM magnetic ordering which is highly energetically unfavourable for both Co/Nb distributions, the valence band is dominated by strongly hybridized Co-3d/O-2p states, while Nb-4d contributions remain negligible, reflecting the ionic Nb$^{5+}$ character. Both rock-salt and layered NM configuration has direct band gaps at $\Gamma$ point (see Fig. \ref{ACBN0_bands}) with values of $E_g\approx2.19$ eV and $E_g\approx1.23$ eV, respectively. 

 The layered-AFM configuration displays sharper, more atomic-like Co-3d features in the valence band and a pronounced suppression of DOS near the Fermi level, showing enhanced electron localization and stronger AFM exchange interactions. The calculated band gap values for the layered AFM structure $E_g^{dir} \approx 0.56$ eV ($A\rightarrow A$) and $E_g^{indir} \approx 0.50$ eV ($\Gamma \rightarrow A$). In contrast, the rock-salt (uniform) AFM configuration exhibits broader Co-3d-derived bands, indicating the enhanced three-dimensional Co-O-Co connectivity compared to the layered arrangement. The calculated band gap values for the rock-salt AFM structure are $E_g^{dir} \approx 0.88$ eV ($Z\rightarrow Z$) and $E_g^{indir} \approx 0.73$ eV ($\Gamma \rightarrow Z$) (Fig. \ref{ACBN0_bands}).
 
 For the layered-FM structure both spin states (IS and HS) are metallic within our approximations, however, for the HS the total density of states at the Fermi level is $\approx 2.5$ times lower than for the IS state. The coexistence of metallic character with a small energy difference between spin states shows the itinerant character of magnetism for the ferromagnetically ordered structure with layered Co/Nb distribution. The rock-salt FM structure has a high-spin Co$^{3+}$ state, and band gaps of $E_g^{dir}\approx 1.07$ eV ($A \rightarrow A$) and  $E_g^{indir}\approx 0.70$ eV ($\Gamma \rightarrow A$)

Across all configurations, the dominance of Co-3d/O-2p hybridization in the valence band and the absence of Nb-4d states near Fermi level confirm that charge transport proceeds almost exclusively via Co-O-Co pathways. The systematic reduction of band gap values upon transitioning from rock-salt to layered cation distribution, and from NM to AFM magnetic order, provides a first-principles explanation for the experimentally observed small-polaron hopping conduction and the high positive Seebeck coefficient. 

The results demonstrate that B-site cation distribution critically affect the physical parameters - bandwidth, hybridization strength, and degree of electron localization.

\begin{table}[width=0.6\linewidth,cols=4,pos=h]
\caption{Converged U values from FHI-aims calculations with ACBN0@PBE and Mulliken projection function for different atoms in rock-salt (uniform) and layered  distribution of Co/Nb ions and different magnetic ordering in Ca$_2$CoNbO$_6$ structure}
\label{U_ACBN0}
\centering
    \begin{tabular*}{\tblwidth}{cccc|cccc}
\toprule
& \multicolumn{3}{c|}{\textbf{rock-salt}} & \multicolumn{4}{c}{\textbf{layered}}\\
\midrule
         & NM & FM & AFM & NM & FM (HS) & FM (IS) & AFM
        \\
        \midrule
        Nb-4d &  0.05 & 0.05 & 0.04 & 0.08 & 0.07 & 0.07 & 0.06\\
        Co-3d & 3.25 & 1.73 & 1.25 & 3.15 & 1.58 & 2.46/2.37  & 1.40\\
        O1-2p & 7.35 & 6.97 & 6.88 & 7.87 & 7.07 & 7.55 & 7.16\\
        O2-2p & 7.35 & 6.99 & 6.89 & 6.48  & 6.75 & 6.33  & 6.56\\
        O3-2p & 7.36 & 7.02 & 6.85 & 7.53 & 6.97 & 7.56  & 6.94\\
        \bottomrule
    \end{tabular*}
\end{table}

\printcredits

\bibliographystyle{model1-num-names}

\bibliography{cas-refs}



\end{document}